\documentclass[journal]{IEEEtran}

\usepackage{cite}
\usepackage{textcomp}
\usepackage{amsmath, amssymb, amsfonts}
\usepackage{algorithm}
\usepackage{algorithmic}
\usepackage{color}
\usepackage{graphicx}
\usepackage{enumerate}
\usepackage{booktabs}
\usepackage{caption}
\usepackage{subcaption}
\usepackage{enumerate}
\usepackage{mathrsfs}
\usepackage{mathtools}
\usepackage{chngcntr}

\makeatletter
\let\NAT@parse\undefined
\makeatother
\usepackage{hyperref}  
\hypersetup{
    colorlinks= true,
    citecolor= black,
    linkcolor= black, 
    pdfborder= {0 0 0},
}

\DeclareMathOperator{\minimize}{minimize}

\DeclareMathOperator{\E}{\mathsf{E}}
\DeclareMathOperator{\Cov}{\mathsf{cov}}

\DeclareMathOperator{\argmin}{arg min}

\newtheorem{definition}{Definition}
\newtheorem{lemma}{Lemma}
\newtheorem{theorem}{Theorem}
\newtheorem{proposition}{Proposition}
\newtheorem{remark}{Remark}
\newtheorem{assumption}{Assumption}

\begin{document}

\title{Stochastic Control with Stale Information--Part I:\\Fully Observable Systems}
\author{
Touraj Soleymani, John S. Baras, and Karl H. Johansson
\thanks{T.~Soleymani, J.~S.~Baras, and K.~H.~Johansson are with the Automatic Control Department, Royal Institute of Technology, SE-11428 Stockholm, Sweden. J.~S.~Baras is also with the Institute for Systems Research, University of Maryland College Park, MD 20742, USA. Email addresses: {\tt \small touraj@kth.se, baras@isr.umd.edu, kallej@kth.se}.}%
}


\maketitle






%
%


\begin{abstract}
In this study, we adopt age of information as a measure of the staleness of information, and take initial steps towards analyzing the control performance of stochastic systems with stale information. Our goals are to cast light on a fundamental limit on the information staleness that is required for a certain level of the control performance and to specify the corresponding stalest information pattern. In the asymptotic regime, such a limit asserts a critical information staleness that is required for stabilization. We achieve these goals by formulating the problem as a stochastic optimization problem and characterizing the associated optimal solutions. These solutions are in fact a control policy, which specifies the control inputs of the plant, and a queuing policy, which specifies the staleness of information at the controller.

\end{abstract}

%

\section{Introduction}
\begin{figure}[t]
  \centering
  \includegraphics[width= 0.85\linewidth]{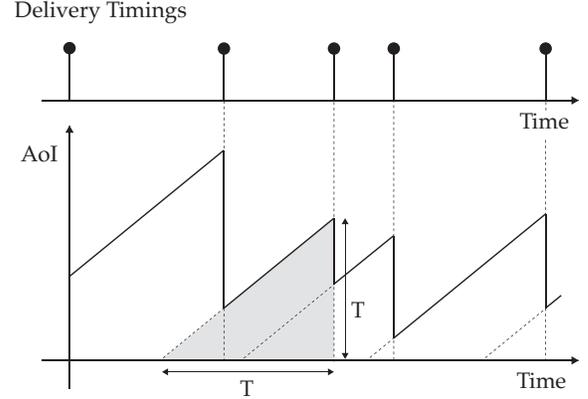}
  \caption{Evolution of the age of information as a function of time with the corresponding delivery timings for a receiver.}
  \label{fig:AoI}
\end{figure}

Age of Information (AoI) has recently received a significant attention in the literature, particularly for systems whose operations depend on time-sensitive information, e.g., monitoring systems that obtain sensory information from the environment. In such systems, a data packet containing too stale information has little value even if it is delivered promptly. The main characteristic of age of information as a measure is that, contrary to delay, it captures the staleness of information from the perspective of the receiver. Formally speaking, age of information (or simply age) measures the time elapsed since the last received update was generated. Fig.~\ref{fig:AoI} schematically shows the evolution of the age as a function of time for a receiver. The age, upon reception of a new update, drops to the time elapsed since this update was generated, and grows linearly otherwise.

In this study, we adopt age of information as a measure of the staleness of information, and take initial steps towards analyzing the control performance of stochastic systems with stale information. Our goals are to cast light on a \emph{fundamental limit} on the information staleness that is required for a certain level of the control performance and to specify the corresponding \emph{stalest information pattern}. In the asymptotic regime, such a limit asserts a \emph{critical} information staleness that is required for stabilization. We achieve these goals by formulating the problem as a stochastic optimization problem and characterizing the associated optimal solutions.

\subsection{Literature Survey on AoI}
The notion of age of information was introduced by Kaul~\emph{et al.}~\cite{kaul2011} in 2011 to quantify the timeliness of information in the problem of congestion control in large wireless networks where nodes periodically broadcast time-sensitive information. They showed that age in a network is in fact minimized at an optimal operating point that lies between the extremes of maximum throughput and minimum delay. Age of information have been adopted in various applications since then, and accordingly different related metrics such as \emph{average age of information} and \emph{peak age of information} have been proposed, see e.g., \cite{kaul2012, yates2016,  kosta2017, talak2018, baknina2018, kadota2018, talak2017, heminimizing, sun2017, jiang2018, mayekar2018}. The importance of age of information is due to the fact that timeliness, which is characterized by age of information, is an emerging requirement for cyber-physical systems~\cite{baheti2011}. However, timeliness has not yet studied systematically from the control perspective.

\subsection{Contributions and Outline}
In Part I of our study, we concentrate on fully observable systems. In Part II, which will be published elsewhere, we will address partially observable systems. The main contributions of the current paper are as follows:
\begin{enumerate}
  \item Exploiting a measure for the staleness of information in stochastic control systems,
  \item Scrutinizing the trade-off between the information staleness and control performance,
  \item Characterizing the stalest information pattern that guarantees a specific control performance.
\end{enumerate}

This paper is organized in the following way. We formally state the problem in Section~\ref{sec:form}. We provide the main results in Section~\ref{sec:main}. Then, we present the numerical results in Section~\ref{sec:examp}. Finally, we conclude the paper in Section~\ref{sec:conc}.



\subsection{Notations}
In the sequel, vectors, matrices, and sets are represented by lower case, upper case, and Calligraphic letters like $x$, $X$, and $\mathcal{X}$ respectively. The sequence of all vectors $x_{t}, \ t=0,\dots,k$, is represented by $\mathbf{x}_k$, and the sequence of all vectors $x_{t}, \ t=k,\dots,N$ for a specific time horizon $N$, is represented by $\mathbf{x}^k$. For matrices $X$ and $Y$, the relations $X \succ 0$ and $Y \succeq 0$ denote that $X$ and $Y$ are positive definite and positive semi-definite respectively. The expected value and covariance of the random variable $x$ are represented by $\E[x]$ and $\Cov[x]$ respectively.



%
%


\begin{figure}[t]
  \centering
  \includegraphics[width= 0.98\linewidth]{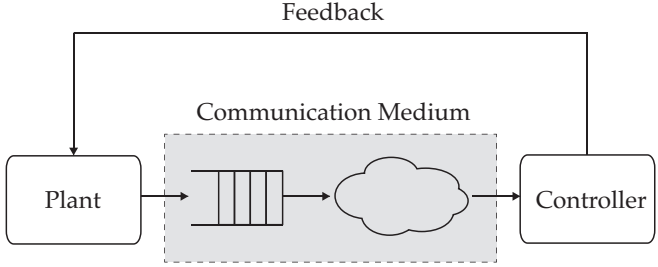}
  \caption{A control system with queuing mechanism in the communication channel between the plant and controller.}
  \label{fig:buffer}
\end{figure}
 
\section{Problem Formulation}\label{sec:form}
Consider a plant, which is to be controlled, with stochastic dynamics governed by the following linear discrete-time state system:
\begin{align}\label{eq:sys}
	x_{k+1} &= A x_k + B u_k + w_k,
\end{align}
with initial condition $x_0$ where $x_k \in \mathbb{R}^n$ is the state of the system, $A_k \in \mathbb{R}^{n \times n}$ is the state matrix, $B_k \in \mathbb{R}^{n \times m}$ is the input matrix, $u_k \in \mathbb{R}^m$ is the control input, and $w_k \in \mathbb{R}^n$ is a Gaussian noise with zero mean and covariance $W_k \succ 0$. It is assumed that the initial state $x_0$ is a Gaussian vector with mean $m_0$ and covariance $M_0$, that $x_0$ and $w_k$ are mutually independent for all $k$, and $(A,B)$ is controllable.

The plant is fully observable, and its measurements are to be transmitted to a controller over a noiseless communication channel. However, there exists a buffer at the transmitter that stores the measurements in a queue, and the priority for transmission is specified by a \emph{queuing mechanism} (see Fig.~\ref{fig:buffer}). The next definitions are useful in our following discussion about the queuing mechanism.

\begin{definition}
	A measurement $x_t$ is said to be \emph{informative} if the latest measurement at the controller is $x_{t'}$ such that $t' < t$.
\end{definition}

\begin{definition}
	A measurement $x_t$ is said to be \emph{obsolete} if there is at least one measurement $x_{t'}$ that is transmitted to the controller such that $t' \geq t$.
\end{definition}

Clearly, a measurement that is not informative is obsolete. At each time $k$, it is possible to transmit one of the measurements in the queue, which are all informative for now, or to transmit no measurement. If a measurement $x_{k'}$ is transmitted, then all the measurements that became obsolete are discarded in the queue. From the resource allocation perspective, the preference is not to transmit or transmit a measurement that arrived earlier in the queue. The waiting time of an informative measurement $x_{k'}$ at time $k$ is equal to $k-k'$.

As a result, the staleness of information at the controller directly depends on the operation of the queuing mechanism. We specify the latest measurement at the controller at time $k$ by $x_{k-\eta_k}$ where $\eta_k$ is called \emph{age of information} at time $k$. In fact, age is an integer such that $\eta_k \in [0, \eta_{k-1} + 1]$. We assume that $\eta_0 = 0$, which means that the controller has access to $x_0$ at time $k=0$. Apart from the case $\eta_k = \eta_{k-1} + 1$, which occurs when no measurement is transmitted, for any $\eta_k \leq \eta_{k-1}$ a measurement is delivered to the controller and the age is equal to the waited time of that measurement in the queue. By convention, we say that the waiting time is equal to $\eta_{k-1} + 1$ if no measurement is transmitted. 

Let $\mu = \{u_0, \dots, u_N \}$ and $\pi = \{\eta_0, \dots, \eta_N\}$ denote a control policy and a queuing policy respectively. We measure the information staleness available at the controller by the expected average age of information expressed~by
\begin{align}
	A(\pi, \mu) =  \textstyle \frac{1}{N} \E \Big[  \sum_{k=0}^{N} \check{\theta}_k \eta_k \Big],
\end{align}
where $\check{\theta}_k$ is a weighting coefficient, and measure the control performance by the expected average quadratic terms of the state and control expressed~by
\begin{equation}
\begin{aligned}
	J(\pi,\mu) =  \textstyle \frac{1}{N} &\E \Big[ x_{N+1}^T Q_{N+1} x_{N+1}\\[1\jot]
	&\quad + \textstyle \sum_{k=0}^{N} x_k^T Q_k x_k + u_k^T R_k u_k  \Big],
\end{aligned}
\end{equation}
where $Q_k \succeq 0$ and $R_k \succ 0$ are weighting matrices. 

Our goal is to obtain the maximum information staleness such that the control performance is guaranteed to be less than or equal to $J_0$. In particular, we seek to find
\begin{align}\label{eq:optimal-age-1}
	A^*(J_0) = \sup_{\pi,\mu: J(\pi,\mu) \leq J_0} A(\pi,\mu).
\end{align}
Following the convexity of the function $A^*(J_0)$, we can introduce the Lagrange multiplier $\lambda \geq 0$ and reformulate~(\ref{eq:optimal-age-1})~as
\begin{align}
	A^*(J_0) = \underset{\pi,\mu}{\inf} \Big \{ - A(\pi,\mu) + \lambda J(\pi,\mu) \Big \} - \lambda J_0.
\end{align}
Hence, in order to find $A^*(J_0)$, we should solve the following stochastic optimization problem
\begin{align}\label{eq:opt-main}
\minimize \quad \chi(\pi,\mu),
\end{align}
over policies $\pi$ and $\mu$ where 
\begin{equation}
\begin{aligned}
	\chi(\pi,\mu) =  \textstyle &\E \Big[ x_{N+1}^T Q_{N+1} x_{N+1}\\[1\jot]
	&\quad + \textstyle \sum_{k=0}^{N} - \theta_k \eta_k + x_k^T Q_k x_k + u_k^T R_k u_k  \Big],
\end{aligned}
\end{equation}
where $\theta_k = \check{\theta}_k/ \lambda$.

The stochastic optimization problem in (\ref{eq:opt-main}) is in general NP-hard. However, we can study this problem under \emph{restricted information sets}, which are described by the following assumptions.

\begin{assumption}\label{assump1}
The information set of the controller is restricted to the information given by the delivered measurements $x_{t - \eta_t}$ for all $t \leq k$, and the information that $\eta_{t'}$ might have about $x_t$ for all $ t \leq t' \leq k$ is neglected.
\end{assumption}

\begin{assumption}\label{assump2}
The information set of the queuing mechanism is restricted to the information that is not predictable by the controller, and the information that is predictable by the controller is neglected.
\end{assumption}

We represent the information set of the controller satisfying Assumption~\ref{assump1} by  $\mathcal{I}^{c}_k = \{ x_{t-\eta_t} | t \leq k  \}$, and the information set of the queuing mechanism satisfying Assumption~\ref{assump2} by $\mathcal{I}^{q}_k = \{ x_t - \tilde{x}_t | t \leq k\}$ where $\tilde{x}_t$ is the best estimate of $x_t$ given $\mathcal{I}^{c}_k$. Then, a control policy $\pi$ is admissible if $u_k$ is a measurable function of $\mathcal{I}^c_k$ for all $k$, and a queuing policy $\pi$ is admissible if $\eta_k$ is a measurable function of $\mathcal{I}^q_k$ for all $k$. In the sequel, we study the solutions of the stochastic optimization problem in (\ref{eq:opt-main}) in the domain of admissible policies.


\section{Main Results}\label{sec:main}
We provide the main results in this section. In particular, we first construct the optimal estimator at the controller. Then, we characterize the optimal control policy and optimal queuing policy associated with the stochastic optimization problem in~(\ref{eq:opt-main}). Finally, we present a suboptimal queuing policy that can efficiently be implemented.

Due to the existence of a queuing mechanism at the transmitter, the controller does not have access to the current state of the plant at each time, and hence requires to employ an estimator. The next proposition provides the optimal estimator at the controller.

\begin{proposition}\label{prop:1}
The optimal estimator minimizing the mean-square error at the controller is given by
\begin{align}
	\hat{x}_k = A^{\eta_k} x_{k-\eta_k} + \textstyle \sum_{t=1}^{\eta_k} A^{t-1} B u_{k-t},
\end{align}
where $\hat{x}_k = \E[x_k | \mathcal{I}^c_k]$ and $\eta_k$ is the age at time $k$.
\end{proposition}
\begin{IEEEproof}
	First of all, it is clear that given the information set $\mathcal{I}^c_k$, the conditional expectation $\hat{x}_k = \E[x_k | \mathcal{I}^c_k]$ minimizes the mean-square error. From the definition, $x_{k-\eta_k}$ represents the latest measurement at the controller. Given $\eta_k$, we can write $x_k$ in terms of $x_{k-\eta_k}$ as follows:
\begin{align*}
	x_k = A^{\eta_k} x_{k-\eta_k} + \textstyle \sum_{t=1}^{\eta_k} A^{t-1} B u_{k-t} + A^{t-1} w_{k-t},
\end{align*}
Taking conditional expectation with respect to the information set $\mathcal{I}^c_k$, we obtain the result:
\begin{align*}
	\E[ x_k| \mathcal{I}^c_k] &= \E \Big[ A^{\eta_k} x_{k-\eta_k} + \textstyle \sum_{t=1}^{\eta_k} A^{t-1} B u_{k-t}\\[1\jot]
	&\qquad \qquad \qquad \qquad \qquad + A^{t-1} w_{k-t} \Big| \mathcal{I}^c_k \Big]\\[1\jot]
	&= A^{\eta_k} x_{k-\eta_k} + \textstyle \sum_{t=1}^{\eta_k} A^{t-1} B u_{k-t},
\end{align*}
where we used the fact that $x_{k-\eta_k}$ and $u_t$ for $k-\eta_k \leq t \leq k-1$ are $\mathcal{I}^c_k$-measurable and $\E[w_t | \mathcal{I}^c_k] = 0$ for all $k-\eta_k \leq t \leq k - 1$.
\end{IEEEproof}

\begin{remark}
The plant is fully observable, and it is easy to show that at time $k$ the innovations since the last update, i.e., the discrepancies between $x_t$ and $\tilde{x}_t = [x_t | \mathcal{I}_k^c]$ for $k-\eta_{k-1}-1 \leq t \leq k$, are functions of the noise $w_t$. Hence, we can equivalently represent the information set of the queuing mechanism as $\mathcal{I}^q_k = \{ w_t | t \leq k-1 \}$. Notice that the initial condition $x_0$ is known at the controller.
\end{remark}

The next lemma provides an identity for the cost function $\chi(\pi,\mu)$, which will be used later.

\begin{lemma}\label{lem:1}
Define the matrix $S_k \succeq 0$ such that for all $k$ it satisfies
\begin{align}\label{eq:riccati}
		S_k = Q_k + A^T S_{k+1} A - K_k^T(R_k + B^T S_{k+1} B) K_k,
\end{align}
with initial condition $S_{N+1} = Q_{N+1}$ where
\begin{align}\label{eq:k}
	K_k &= (R_k + B^T S_{k+1} B)^{-1} B^T S_{k+1} A.
\end{align}
Then, the cost function $\chi(\pi,\mu)$ is equal to
\begin{equation}
\begin{aligned}
	&\chi(\pi,\mu) = \E \Big[ x_0^T S_0 x_0 + \textstyle \sum_{k=0}^{N} -\theta_k \eta_k +  w_k^T S_{k+1} w_k\\[1\jot]
	&\qquad + (u_k + K_k x_k)^T (B^T S_{k+1} B + R_k) (u_k + K_k x_k)\Big].	\\[1\jot]
\end{aligned}
\end{equation}
\begin{IEEEproof}
Using (\ref{eq:sys}) and (\ref{eq:riccati}), we have
\begin{align}
\begin{split}\label{eq:sk1}
	x_{k+1}^T S_{k+1} x_{k+1} &= (A x_k + B u_k + w_k)^T\\[2\jot]
	&\qquad \times S_{k+1} 	(A x_k + B u_k + w_k),
\end{split}\\[2\jot]
\begin{split}\label{eq:sk}
	x_k^T S_k x_k &= x_k^T \big(Q_k + A^T S_{k+1} A\\[2\jot]
	&\qquad - K_k ^T (B^T S_{k+1} B + R_k) K_k\big) x_k.
\end{split}
\end{align}
Moreover, we can write
\begin{align*}
	&x_{N+1}^T S_{N+1} x_{N+1} - x_0^T S_0 x_0\\[2\jot]
	&\qquad  =  \textstyle \sum_{k=0}^{N} x_{k+1}^T S_{k+1} x_{k+1} - x_k^T S_k x_k\\[2\jot]
	&\qquad  =  \textstyle \sum_{k=0}^{N} w_k^T S_{k+1} w_k + 2 (A x_k + B u_k)^TS_{k+1} w_k \\[2\jot]
	&\qquad \quad + x_k^T K_k ^T (B^T S_{k+1} B + R_k) K_k x_k\\[2\jot]
	&\qquad \quad - x_k^T Q_k x_k - u_k^T R_k u_k + 2 x_k^T A^T S_{k+1} B u_k\\[2\jot]
	&\qquad \quad + u_k^T (B^T S_{k+1} B + R_k) u_k,
\end{align*}
where the first equality is an identity, and in the second equality we used (\ref{eq:sk1}) and (\ref{eq:sk}) and also added and subtracted the term $\sum_{k=0}^{N} u_k^T R_k u_k$ to and from the right-hand side. Rearranging the terms in the above relation, we find
\begin{equation*}
\begin{aligned}
	& x_{N+1}^T S_{N+1} x_{N+1} + \textstyle \sum_{k=0}^{N} x_k^T Q_k x_k + u^T R_k u_k\\[2\jot]
	& = x_0^T S_0 x_0 + \textstyle \sum_{k=0}^{N} w_k^T S_{k+1} w_k + 2 (A x_k + B u_k)^TS_{k+1} w_k\\[2\jot]
	&\quad + (u_k + K_k x_k)^T (B^T S_{k+1} B + R_k) (u_k + K_k x_k).
\end{aligned}
\end{equation*}
Adding the term $\sum_{k=0}^{N} - \theta_k \eta_k$ to the both sides of the above relation and taking expectation, we obtain the result:
\begin{align*}
	\chi(\pi,\mu) &= \E \Big[ x_0^T S_0 x_0 + \textstyle \sum_{k=0}^{N} -\theta_k \eta_k + w_k^T S_{k+1} w_k\\[2\jot]
	&\quad + 2(A x_k + B u_k)^T S_{k+1} w_k\\[1\jot]
	&\quad + (u_k + K_k x_k)^T (B^T S_{k+1} B + R_k) (u_k + K_k x_k) \Big]
\end{align*}	
\begin{align*}
	& = \E \Big[ x_0^T S_0 x_0 + \textstyle \sum_{k=0}^{N} - \theta_k \eta_k +  w_k^T S_{k+1} w_k\\[1\jot]
	&\quad + (u_k + K_k x_k)^T (B^T S_{k+1} B + R_k) (u_k + K_k x_k)\Big],
\end{align*}
where in the second equality we used the fact that $w_k$ is independent of $x_k$ and $u_k$.
\end{IEEEproof}
\end{lemma}

We recall that the control policy $\pi$ is admissible if $u_k$ is a measurable function of $\mathcal{I}^c_k$ for all $k$. The next theorem characterizes the optimal control policy.

\begin{theorem}\label{thm:control-policy}
The optimal control policy is a certainty-equivalence policy given by
\begin{align}
	u_k^* = - K_k \hat{x}_k,
\end{align}
where $K_k$ is defined in (\ref{eq:k}).

\begin{IEEEproof}
From Proposition~\ref{prop:1}, we can obtain the estimation error:
\begin{equation}\label{eq:error-cal}
\begin{aligned}
	e_k &= x_k - \hat{x}_k\\[2\jot]
	&= x_k - A^{\eta_k} x_{k-\eta_k} - \textstyle \sum_{t=1}^{\eta_k} A^{t-1} B u_{k-t} \\[2\jot]	
	&= \textstyle \sum_{t=1}^{\eta_k} A^{t-1} w_{k-t}.
\end{aligned}
\end{equation}
We note that $\eta_k$ is a function of $w_t$ for $k-\eta_{k-1}-1 \leq t \leq k$. Therefore, the estimation error is independent of the control policy.

In addition, using the identity $x_k = \hat{x}_k + e_k$ in the cost function $\chi(\pi,\mu)$ given by Lemma~\ref{lem:1} we obtain
\begin{equation*}
\begin{aligned}
	&\chi(\pi,\mu) = \E \Big[ x_0^T S_0 x_0 + \textstyle \sum_{k=0}^{N} - \theta_k \eta_k +  w_k^T S_{k+1} w_k\\[1\jot]
	&\qquad + (u_k + K_k \hat{x}_k +  K_k e_k)^T (B^T S_{k+1} B + R_k)\\[1\jot]
	&\qquad \times(u_k + K_k \hat{x}_k +  K_k e_k)\Big].
\end{aligned}
\end{equation*}

Following the fact that the terms $x_0^T S_0 x_0$, $\sum_{k=0}^{N} \theta_k \eta_k$, and $\sum_{k=0}^{N}w_k^T S_{k+1} w_k$ are independent of the control policy, we define the value function $V^c_k$ as
\begin{align*}
	V^c_k &= \min_{\mathbf{u}^k}\E \Big[ \textstyle \sum_{t=k}^{N} (u_t + K_t \hat{x}_t +  K_t e_t)^T\\[1\jot]
	&\qquad \times (B^T S_{t+1} B + R_t) (u_t + K_t \hat{x}_k +  K_t e_t) \Big| \mathcal{I}^c_k \Big].
\end{align*}
Consequently, we have
\begin{align*}	
	V^c_k &= \min_{\mathbf{u}^k}\E \Big[ \textstyle \sum_{t=k}^{N} (u_t + K_t \hat{x}_t)^T\\[1\jot]
	&\qquad \times (B^T S_{t+1} B + R_t) (u_t + K_t \hat{x}_t)\\[2\jot]
	&\qquad + e_t^T K_t^T (B^T S_{t+1} B + R_t) K_t e_t\\[1\jot]
	&\qquad + 2(u_t + K_t \hat{x}_t)^T (B^T S_{t+1} B + R_t) K_t e_t \Big| \mathcal{I}^c_k \Big]\\[1\jot]
	&=\min_{\mathbf{u}^k} \Big\{ \textstyle \sum_{t=k}^{N} (u_t + K_t \hat{x}_t)^T \\[1\jot]
	&\qquad \times (B^T S_{t+1} B + R_t) (u_t + K_t \hat{x}_t)\\[1\jot]
	&\qquad + \E \big[ e_t^T K_t^T (B^T S_{t+1} B + R_t) K_t e_t \big| \mathcal{I}^c_k \big] \Big\},
\end{align*}
where in the second equality we used the fact that $u_k$ and $\hat{x}_k$ are $\mathcal{I}^c_k$-measurable and $\E[ e_k | \mathcal{I}^c_k] = 0$. Now, we can conclude that $u^*_k = -K_k \hat{x}_k$. This completes the proof.	
\end{IEEEproof}
\end{theorem}

\begin{figure}[t]
  \centering
  \includegraphics[width= 0.98\linewidth]{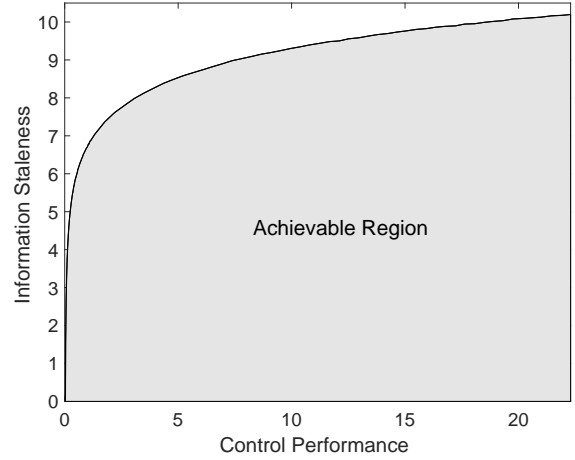}
  \caption{Trade-off curve between the information staleness and control performance. The control performance is scaled by one thousandth. The area under the trade-off curve represents the achievable region of stochastic control with stale information.}
  \label{fig:tradeoff}
\end{figure}

We recall that the queuing policy $\pi$ is admissible if $\eta_k$ is a measurable function of $\mathcal{I}^q_k$ for all $k$. The following theorem characterizes the optimal queuing policy.

\begin{theorem}\label{thm:queuing-policy}
The optimal queuing policy is given by
\begin{equation}
\begin{aligned}
	\eta_k^* &= \underset{\eta_k \in [0,\eta_{k-1}+1]}{\argmin} \Big \{ - \theta_k \eta_k + c_k\\[1\jot]
	&\quad + \textstyle \big(\sum_{t=1}^{\eta_k} A^{t-1} w_{k-t} \big)^T \Gamma_k \big(\sum_{t=1}^{\eta_k} A^{t-1} w_{k-t} \big) \Big \},
\end{aligned}
\end{equation}
where $c_k$ is a variable that depends on $\eta_k$ and $\Gamma_k = K_k^T (B^T S_{k+1} B + R_k) K_k$.

\begin{IEEEproof}
Incorporating the optimal control policy $u^*_k = -K_k \hat{x}_k$ in the cost function $\chi(\pi,\mu)$ given by Lemma~\ref{lem:1}, we obtain
\begin{align*}
	\chi(\pi,\mu) & = \E \Big[ x_0^T S_0 x_0 + \textstyle \sum_{k=0}^{N} - \theta_k \eta_k + w_k^T S_{k+1} w_k\\[2\jot]
	&\qquad \qquad + e_k^T K_k^T (B^T S_{k+1} B + R_k) K_k e_k  \Big].
\end{align*}

Following the fact that the terms $x_0^T S_0 x_0$ and $\sum_{k=0}^{N}w_k^T S_{k+1} w_k$ are independent of the queuing policy, we define the value function $V^q_k$ as
\begin{align*}
	V^q_k &= \min_{\boldsymbol{\eta}^k} \E \Big[ \textstyle \sum_{t=k}^{N} - \theta_t \eta_t + e_t^T \Gamma_t e_t \Big| \mathcal{I}^q_k \Big].
\end{align*}
Consequently, we have
\begin{align*}		
	V^q_k &= \min_{\eta_k} \E \Big[ - \theta_k \eta_k + e_k^T \Gamma_k e_k + V^q_{k+1} \Big| \mathcal{I}^q_k \Big]\\[1\jot]
	&=\min_{\eta_k} \Big\{ - \theta_k \eta_k + e_k^T \Gamma_k e_k + \E[V^q_{k+1} | \mathcal{I}^q_k] \Big\},
\end{align*}	
where in the first equality we used the linearity of the value function $V^q_k$ with $V^q_{N+1} = 0$ and in the second equality the fact that $e_k$ is $\mathcal{I}^q_k$-measurable. The result is obtained by defining $c_k = \E[V^q_{k+1} | \mathcal{I}^q_k]$ and using (\ref{eq:error-cal}).
\end{IEEEproof}
\end{theorem}

\begin{figure}[t]
  \centering
  \includegraphics[width= 0.98\linewidth]{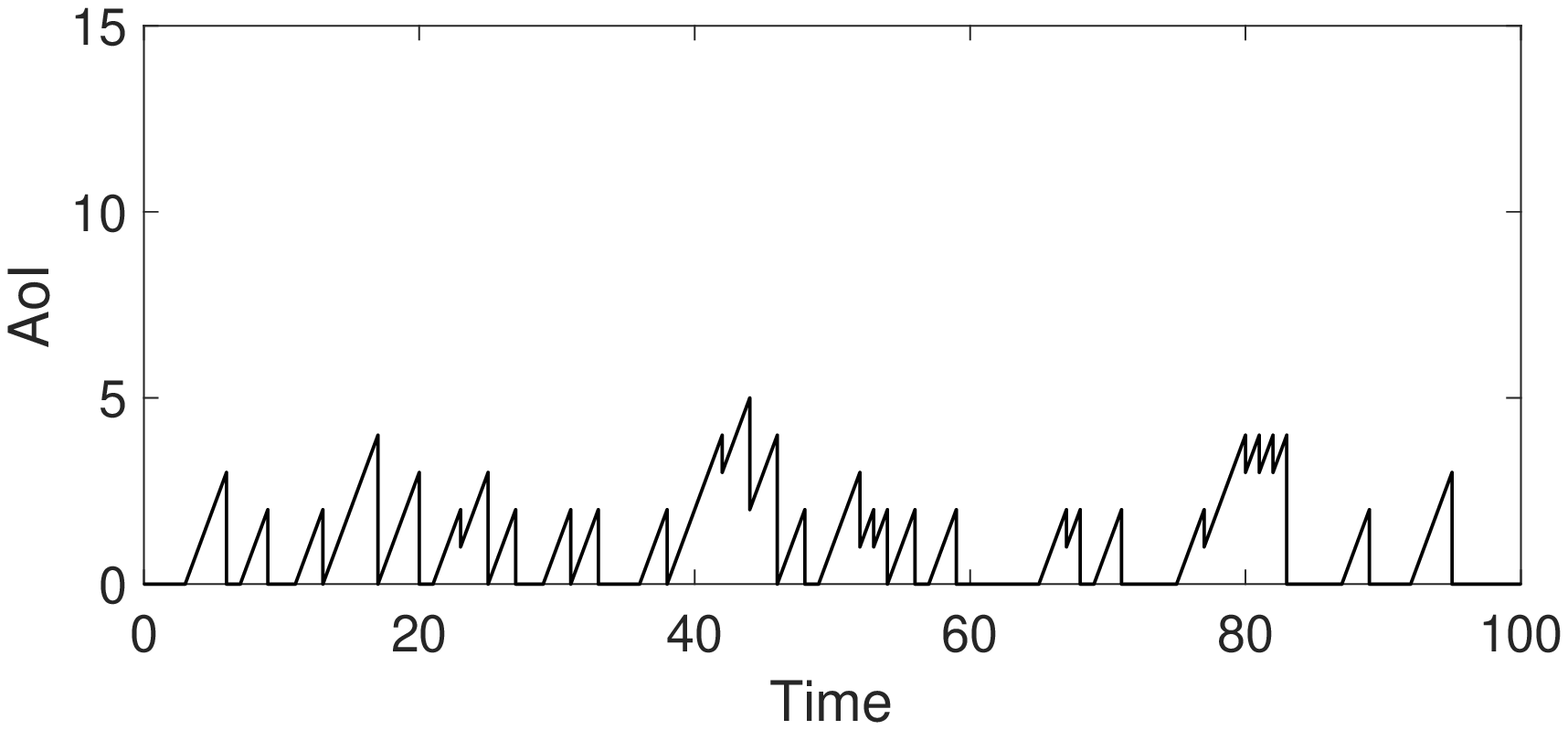}
  \caption{Evolution of the age of information for $\lambda = 0.1$.}
  \label{fig:aoi-1}
\end{figure}

\begin{figure}[t]
  \centering
  \includegraphics[width= 0.98\linewidth]{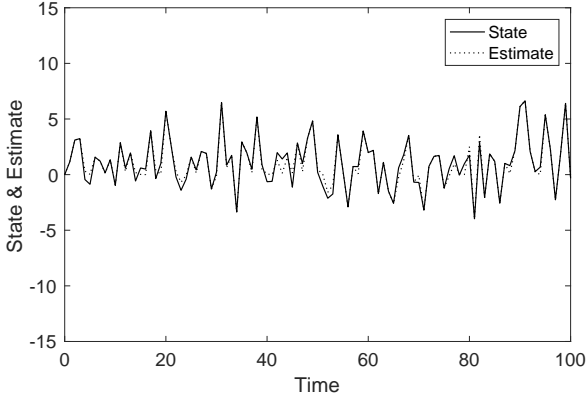}
  \caption{Trajectories of the state and estimate for $\lambda = 0.1$.}
  \label{fig:state-1}
\end{figure}

The following proposition provides a suboptimal queuing policy with a performance guarantee that can efficiently be implemented.

\begin{proposition}\label{prop:subopt}
A suboptimal queuing policy	that outperforms the zero-wait queuing policy (i.e., the queuing policy with $\eta_k = 0$ for all $k$) is given by
\begin{equation}\label{eq:suboptimal-policy-prop}
\begin{aligned}
	\eta_k^+ &= \underset{\eta_k \in [0,\eta_{k-1}+1]}{\argmin} \Big \{ - \theta_k \eta_k\\[1\jot]
	&\quad + \textstyle \big(\sum_{t=1}^{\eta_k} A^{t-1} w_{k-t}\big)^T \Gamma_k \big(\sum_{t=1}^{\eta_k} A^{t-1} w_{k-t}\big) \Big \},
\end{aligned}
\end{equation}
where $\Gamma_k = K_k^T (B^T S_{k+1} B + R_k) K_k$.
\begin{IEEEproof}
Let $\pi'$ denote the suboptimal queuing policy given by (\ref{eq:suboptimal-policy-prop}) and $\pi''$ denote the zero-wait queuing policy. Clearly, in order to prove that $\chi(\pi',\mu^*) \leq \chi(\pi'',\mu^*)$ it is enough to show that $V^{q'}_{k} \leq V^{q''}_{k}$. This holds at time $N+1$ as $V^{q'}_{N+1} = V^{q''}_{N+1} = 0$. Assume that the claim holds at time $k+1$, we prove that it also holds at time $k$. We have
\begin{align*}		
	V^{q'}_k &= - \theta_k \eta'_k + {e_k'}^T \Gamma_k e'_k  + \E[V^{q'}_{k+1} | \mathcal{I}^q_k] \\[2\jot]
	&\leq - \theta_k \eta'_k + {e_k'}^T \Gamma_k e'_k + \E[V^{q''}_{k+1} | \mathcal{I}^q_k] \\[2\jot]
	&\leq - \theta_k \eta''_k + {e_k''}^T \Gamma_k e''_k + \E[V^{q''}_{k+1} | \mathcal{I}^q_k]  = V^{q''}_k,
\end{align*}
where $e'_k$ and $e''_k$ are the estimation errors associated with $\eta_k'$ and $\eta_k''$ respectively. Moreover, for the zero-wait queuing policy we have
\begin{align*}
	\E[ V^{q''}_{k+1} | \mathcal{I}^q_k]  = \E \Big[ \textstyle \sum_{t=k+1}^{N} - \theta_t \eta''_t +  {e_t''}^T \Gamma_t e''_t \Big| \eta''_t = 0, \mathcal{I}^q_k \Big] = 0.
\end{align*}
This completes the proof.
\end{IEEEproof}
\end{proposition}

\begin{remark}
	One may implement the suboptimal queuing policy in Proposition~\ref{prop:subopt}, with some degradation in the performance, by imposing a limited memory $\bar{k}$ for the queue. In this case, the argument in (\ref{eq:suboptimal-policy-prop}) changes to $\eta_k \in [0, \bar{k}]$.
\end{remark}

\section{Example}\label{sec:examp}
In this section, we use a very simple example to show how the system we developed works essentially. Consider the following scalar stochastic dynamics:
\begin{align}
	x_{k+1} = 1.5 x_{k} + 0.5 u_k + w_k,
\end{align}
with initial condition $x_0 = 0$ and noise variance $W_k = 4$ for all $k$. The weighting coefficients are $Q_k = 5$, $R_k = 0.1$ for all $k$, $Q_{N+1} = 10$, and $\check{\theta}_k = 1$. The time horizon is $N = 100$. Following Proposition~1, the optimal estimator is
\begin{align}
\hat{x}_k = (1.5)^{\eta_k} x_{k-\eta_k} + \textstyle \sum_{t=1}^{\eta_k} (1.5) ^{t-1} 0.5 u_{k-t}	
\end{align}
For this system, we obtained the optimal control policy and a suboptimal queuing policy based on Theorem~\ref{thm:control-policy} and Proposition~\ref{prop:subopt} respectively.

\begin{figure}[t]
  \centering
  \includegraphics[width= 0.98\linewidth]{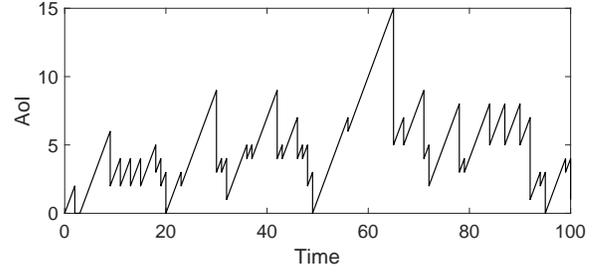}
  \caption{Evolution of the age of information for $\lambda = 0.01$.}
  \label{fig:aoi-50}
\end{figure}

\begin{figure}[t]
  \centering
  \includegraphics[width= 0.98\linewidth]{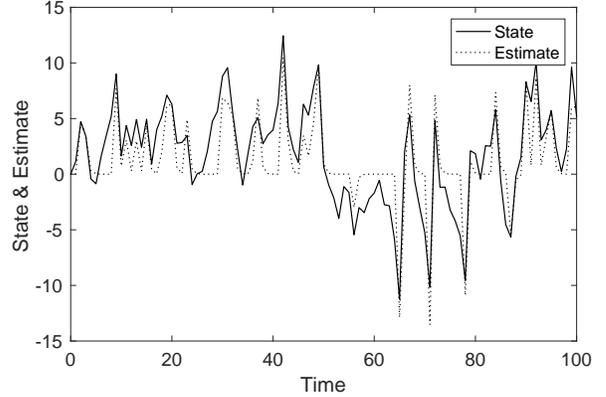}
  \caption{Trajectories of the state and estimate for $\lambda = 0.01$.}
  \label{fig:state-50}
\end{figure}

The trade-off curve between the information staleness and control performance is depicted in Fig.~\ref{fig:tradeoff} for different values of $\lambda$. Note that since we used restricted information sets and a suboptimal queuing policy the obtained trade-off curve should be regarded as a lower bound on $A^*$. The area under the trade-off curve represents the achievable region of stochastic control with stale information. It is interesting to look at the asymptotic behavior of the trade-off curve, which implies that there exists a critical information staleness above which the system is not stabilizable. This asymptotic bound on the information staleness should be studied further in future research. We conjecture that this bound is correlated to the minimum rate required for stabilization of a stochastic system (see e.g., \cite{nair2004}).

In addition, we carried out simulation experiments for two different values of the Lagrange multiplier: $\lambda = 0.1$ and $\lambda = 0.01$. In both cases, we used the same realizations of the noise. The evolution of the age of information and the trajectories of the state and estimate for $\lambda = 0.1$ are shown in Fig.~\ref{fig:aoi-1} and Fig.~\ref{fig:state-1} respectively, and for $\lambda = 0.01$ are shown in Fig.~\ref{fig:aoi-50} and Fig.~\ref{fig:state-50} respectively. As it is seen, when $\lambda$ is smaller the age is allowed to attain larger values, which consequntely leads to a degradation in the qualities of estimation and control. 


%
%
%

\section{Conclusion}\label{sec:conc}
In this paper, we studied the trade-off between the information staleness and control performance for fully observable systems. We adopted the notion of age of information for measuring the staleness of information. We formulated the problem as a stochastic optimization problem, and characterized the associated optimal policies. We proved that the optimal control policy is certainty-equivalence and the optimal queuing policy is a function of the estimation errors associated with the informative measurements.

%

%

\bibliography{mybib.bib}
\bibliographystyle{ieeetr}

\end{document}